\documentclass[aps,prd,twocolumn,superscriptaddress]{revtex4}
\usepackage{epsfig,epsf}
\usepackage{amsmath}
\usepackage{amsthm}
\usepackage{amsfonts}
\usepackage{amssymb}
\usepackage{dsfont}
\usepackage{multirow}
\usepackage{appendix}
\usepackage{slashed}
\usepackage[active]{srcltx}
\usepackage{psfrag}

\begin{document}

\title{Near-threshold structures in the $D_{s}^{+}D_{s}^{-}$ mass
distribution of the decay $B^{+}\rightarrow D_{s}^{+}D_{s}^{-}K^{+}$ }
\date{\today}
\author{S.~S.~Agaev}
\affiliation{Institute for Physical Problems, Baku State University, Az--1148 Baku,
Azerbaijan}
\author{K.~Azizi}
\affiliation{Department of Physics, University of Tehran, North Karegar Avenue, Tehran
14395-547, Iran}
\affiliation{Department of Physics, Do\v{g}u\c{s} University, Dudullu-\"{U}mraniye, 34775
Istanbul, Turkey}
\author{H.~Sundu}
\affiliation{Department of Physics Engineering, Istanbul Medeniyet University, 34700
Istanbul, Turkey}
\affiliation{Department of Physics, Kocaeli University, 41380 Izmit, Turkey}

\begin{abstract}
Two near-threshold peaking structures with spin-parities $J^{\mathrm{PC}%
}=0^{++}$ were recently discovered by the LHCb Collaboration in the $%
D_{s}^{+}D_{s}^{-}$ invariant mass distribution of the decay $%
B^{+}\rightarrow D_{s}^{+}D_{s}^{-}K^{+}$. The first of them is the
resonance $X(3960)$, whereas the second one, $X_0(4140)$, is a structure
with the mass around $4140~\mathrm{MeV}$. To explore their natures and model
them, we study the hadronic molecule $\mathcal{M}=D_s^{+}D_s^{-}$ and
calculate its mass, current coupling, and width. The mass and current
coupling of the molecule are extracted from the QCD two-point sum rule
analyses by taking into account vacuum condensates up to dimension $10$. To
evaluate its full width, we consider the processes $\mathcal{M} \to
D_{s}^{+}D_{s}^{-}$, $\mathcal{M} \to\eta_{c}\eta^{(\prime)}$, and $\mathcal{%
M} \to J/\psi\phi$. Partial widths of these decays are determined by the
strong couplings $g_i, \ i=1,2,3,4 $ at vertices $\mathcal{M}%
D_{s}^{+}D_{s}^{-}$, $\mathcal{M}\eta_{c} \eta^{(\prime)}$, and $\mathcal{M}
J/\psi\phi$. They are computed by means of the three-point sum rule method.
Predictions for the mass $m=(4117 \pm 85)~\mathrm{MeV}$ and width $\Gamma_{%
\mathcal{M}}=(62 \pm 12)~\mathrm{MeV}$ of the molecule $\mathcal{M}$ are
compared with the corresponding LHCb data, and also with our results for the
diquark-antidiquark state $X=[cs][\overline{c}\overline{s}]$. We argue that
the structure $X_0(4140)$ may be interpreted as the hadronic molecule $%
D_s^{+}D_s^{-}$, whereas the resonance $X(3960)$ can be identified with the
tetraquark $X$.
\end{abstract}

\maketitle


\section{Introduction}

\label{sec:Int}

Different $X$ resonances discovered and studied during past years by the
LHCb Collaboration became important part of exotic hadron spectroscopy.
Thus, the resonances $X(4140)$, $X(4274)$, $X(4500)$ and $X(4700)$ were seen
in the process $B^{+}\rightarrow J/\psi \phi K^{+}$ as peaks in the $J/\psi
\phi $ invariant mass distribution \cite{Aaij:2016iza}. The structures $%
X(4140)$ and $X(4274)$ bear the quantum numbers $J^{\mathrm{PC}}=1^{++}$,
and in the tetraquark model are composed of the quarks $c\overline{c}s%
\overline{s}$. The resonances $X(4500)$ and $X(4700)$ are scalar particles
with spin-parities $J^{\mathrm{PC}}=0^{++}$ and the same quark content. It
is worth noting that $X(4140)$ and $X(4274)$ were observed previously in the
decays $B^{\pm }\rightarrow J/\psi \phi K^{\pm }$ by different
collaborations \cite{Aaltonen:2009tz,Chatrchyan:2013dma,Abazov:2013xda},
whereas the scalar resonances $X(4500)$ and $X(4700)$ were discovered for
the first time by LHCb. The resonance $X(4630)$ fixed in the $J/\psi \phi $
invariant mass distribution of the decay $B^{+}\rightarrow J/\psi \phi K^{+}$
is a vector member of the $X$ tetraquarks' family \cite{LHCb:2021uow}.

Recently, LHCb reported new hidden charm-strange structures in the $%
D_{s}^{+}D_{s}^{-}$ invariant mass distribution of the decay $%
B^{+}\rightarrow D_{s}^{+}D_{s}^{-}K^{+}$ \cite{LHCb:2022vsv}. One of them $%
X(3960)$ is presumably a tetraquark $c\overline{c}s\overline{s}$ with
quantum numbers $J^{\mathrm{PC}}=0^{++}$, and the following parameters:
\begin{eqnarray}
m_{1\mathrm{exp}} &=&(3956\pm 5\pm 10)~\mathrm{MeV},  \notag \\
\Gamma _{1\mathrm{exp}} &=&(43\pm 13\pm 8)~\mathrm{MeV}.  \label{eq:Data}
\end{eqnarray}%
This structure is approximately $20~\mathrm{MeV}$ above the $%
D_{s}^{+}D_{s}^{-}$ threshold. The LHCb also found evidence for a second
structure $X_{0}(4140)$ with the mass around $4140~\mathrm{MeV}$ and higher $%
\sim 17~\mathrm{MeV}$ than the $J/\psi \phi $ threshold. The mass and full
width of this state are
\begin{eqnarray}
m_{2\mathrm{exp}} &=&(4133\pm 6\pm 6)~\mathrm{MeV},  \notag \\
\Gamma _{2\mathrm{exp}} &=&(67\pm 17\pm 7)~\mathrm{MeV}.  \label{eq:DataA}
\end{eqnarray}%
The $X_{0}(4140)$ may be interpreted as a new resonance with either a $J^{%
\mathrm{PC}}=0^{++}$ assignment or a $J/\psi \phi \leftrightarrow
D_{s}^{+}D_{s}^{-}$ coupled-channel effect \cite{LHCb:2022vsv}. In the
present work, we assume that the structure $X_{0}(4140)$ is a second
resonance seen by LHCb in the $D_{s}^{+}D_{s}^{-}$ mass distribution.

The $X$ resonances are interesting objects for theoretical investigations:
Features of exotic mesons $c\overline{c}s\overline{s}$ were studied in
numerous articles by employing different models and technical methods\ \cite%
{Nieves:2012tt,Wang:2013exa,Lebed:2016yvr,Chen:2017dpy,Agaev:2017foq,Meng:2020cbk,Sundu:2018toi,Agaev:2022iha}%
. Some of these states have undergone to rather detailed exploration, which
is also provided in our publications. Thus, the axial-vector resonances $%
X(4140)$ and $X(4274)$ were analyzed in Ref.\ \cite{Agaev:2017foq}, in which
they were modeled as states composed of scalar and axial-vector
(anti)diquarks. In the case of $X(4140)$ the constituent diquark
(antidiquark) is the antitriplet (triplet) state of the color group $%
SU_{c}(3)$, whereas to model $X(4274)$ we used (anti)diquarks from the
sextet representation of $SU_{c}(3)$. Predictions for masses and full widths
of these tetraquarks were compared with the LHCb data. It turned out, for
masses of the resonances $X(4140)$ and $X(4274)$ these models led to nice
agreements with the LHCb data. The model based on color triplet
(anti)diquarks also reproduced the full width of $X(4140)$; therefore, it
could be considered as a serious candidate to resonance $X(4140)$. The width
of the construction with color sextet constituents is wider than that of $%
X(4274)$, which excludes it from a list of possible pretenders.

The vector resonances $Y(4660)$ and $X(4630)$ were explored in our articles
(see, Refs. \cite{Sundu:2018toi,Agaev:2022iha}) as tetraquarks $[cs][%
\overline{c}\overline{s}]$ with spin-parities $J^{\mathrm{PC}}=1^{--}$ and $%
J^{\mathrm{PC}}=1^{-+}$, respectively. Predictions for the masses and full
widths of these states allowed us to interpret $Y(4660)$ and $X(4630)$ as
exotic mesons with diquark-antidiquark composition.

The discovery of structures $X(3960)$ and $X_{0}(4140)$ activated
investigations of hidden charm-strange resonances to account for features of
these new states \cite%
{Bayar:2022dqa,Ji:2022uie,Xin:2022bzt,Xie:2022lyw,Chen:2022dad,Guo:2022ggl}.
The $X(3960)$ was considered as a coupled-channel effect \cite{Bayar:2022dqa}%
, or as near the $D_{s}^{+}D_{s}^{-}$ threshold enhancement by the
conventional $P$-wave charmonium $\chi _{c0}(2P)$ \cite{Guo:2022ggl}. The
hadronic $D_{s}^{+}D_{s}^{-}$ molecule model was suggested in Ref.\ \cite%
{Xin:2022bzt} to explain observed properties of the resonance $X(3960)$.

In our paper \cite{Agaev:2022pis}, we examined the tetraquark $X=[cs][%
\overline{c}\overline{s}]$ with quantum numbers $J^{\mathrm{PC}}=0^{++}$ and
calculated its mass and full width. The spectroscopic parameters of $X$,
i.e., its mass and current coupling, were found by means of the QCD
two-point sum rule method. The full width of this state was evaluated using
decay channels $X\rightarrow D_{s}^{+}D_{s}^{-}$ and $X\rightarrow \eta
_{c}\eta ^{(\prime )}$. Predictions for the mass $m=(3976\pm 85)~\mathrm{MeV}
$ and width $\Gamma _{\mathrm{X}}=(42.2\pm 8.3)~\mathrm{MeV}$ obtained in
Ref. \cite{Agaev:2022pis} allowed us to consider the diquark-antidiquark
state $X$ as an acceptable model for $X(3960)$.

In the present article, we continue our studies of the resonance $X(3960)$
and also include in the analysis the structure $X_{0}(4140)$. We investigate
the hadronic molecule $\mathcal{M}=D_{s}^{+}D_{s}^{-}$ by computing its
mass, current coupling, and full width. The mass and current coupling of
this state are calculated in the context of the QCD two-point sum rule
approach. The full width of $\mathcal{M}$ is estimated by considering the
decay channels $\mathcal{M}\rightarrow D_{s}^{+}D_{s}^{-}$, $\mathcal{M}%
\rightarrow \eta _{c}\eta $, $\mathcal{M}\rightarrow \eta _{c}\eta ^{\prime
} $, and $\mathcal{M}\rightarrow J/\psi \phi $. Partial widths of these
processes, apart from parameters of the initial and final particles, depend
also on strong couplings $g_{i},\ i=1,2,3,4$ at vertices $\mathcal{M}%
D_{s}^{+}D_{s}^{-}$, $\mathcal{M}\eta _{c}\eta ^{\prime }$, $\mathcal{M}\eta
_{c}\eta $, and $\mathcal{M}J/\psi \phi $, respectively. To extract
numerical values of $g_{i}$, we use the QCD three-point sum rule method.
Predictions for the mass and width of the molecule $\mathcal{M}$ are
compared with the LHCb data for the resonances $X(3960)$ and $X_{0}(4140)$.
They are also confronted with parameters of the diquark-antidiquark state $%
X=[cs][\overline{c}\overline{s}]$.

This paper is organized in the following way: In Sec.\ \ref{sec:Mass}, we
compute the mass and current coupling of the molecule $\mathcal{M}$ by
employing the QCD two-point sum rule method. The dominant process $\mathcal{M%
}\rightarrow D_{s}^{+}D_{s}^{-}$ is considered in Sec.\ \ref{sec:Decays},
where we determine the coupling $g_{1}$ and partial width of this decay. The
decays $\mathcal{M}\rightarrow \eta _{c}\eta ^{(\prime )}$ and $\mathcal{M}%
\rightarrow J/\psi \phi $ are analyzed in Sec.\ \ref{sec:DecaysA}. The full
width of $\mathcal{M}$ is evaluated also in this section. Section\ \ref%
{sec:Conclusion} is reserved for discussion and summing up.


\section{Spectroscopic parameters of the molecule $\mathcal{M}%
=D_{s}^{+}D_{s}^{-}$}

\label{sec:Mass}

The mass and current coupling of the molecule $\mathcal{M}%
=D_{s}^{+}D_{s}^{-} $ can be evaluated using the QCD two-point sum rule
method \cite{Shifman:1978bx,Shifman:1978by}. This approach works quite well
not only for analysis of conventional particles, but also leads to reliable
predictions in the case of multiquark hadrons.

The principal quantity in this method is an interpolating current for a
hadron under analysis. In the case of the molecule $D_{s}^{+}D_{s}^{-}$,
this current $J(x)$ has a rather simple form
\begin{equation}
J(x)=\left[ \overline{s}_{a}(x)i\gamma _{5}c_{a}(x)\right] \left[ \overline{c%
}_{b}(x)i\gamma _{5}s_{b}(x)\right] ,  \label{eq:C1}
\end{equation}%
where $a$, and $b$ are color indices. This current belongs to $\mathbf{[1}%
_{c}\mathbf{]}_{\overline{s}c}\mathbf{\otimes \lbrack 1}_{c}\mathbf{]}_{%
\overline{c}s}$ representation of the color group $SU_{c}(3)$. It
corresponds to a molecule structure with spin-parities $J^{\mathrm{PC}%
}=0^{++}$, but may also couple to different diquark-antidiquark structures
and other four-quark hadronic molecules \cite{Chen:2022sbf,Xin:2021wcr}.

To find sum rules for the mass $m$ and coupling $f$ of the molecule $%
\mathcal{M}$, one has to start from the calculation of the following
correlation function:
\begin{equation}
\Pi (p)=i\int d^{4}xe^{ipx}\langle 0|\mathcal{T}\{J(x)J^{\dag
}(0)\}|0\rangle .  \label{eq:CF1}
\end{equation}%
At the first stage, we express $\Pi (p)$ in terms of the spectral parameters
of $\mathcal{M}$. To this end, it is necessary to insert a complete set of
states with $J^{\mathrm{PC}}=0^{++}$ into the correlation function $\Pi (p)$
and carry out integration over $x$. These operations lead to the simple
formula
\begin{equation}
\Pi ^{\mathrm{Phys}}(p)=\frac{\langle 0|J|\mathcal{M}(p\rangle \langle
\mathcal{M}(p)|J^{\dagger }|0\rangle }{m^{2}-p^{2}}+\cdots .
\label{eq:PhysSide}
\end{equation}%
The expression derived by this method is a hadronic representation of the
correlator $\Pi (p)$, which forms the phenomenological side of sum rules.
The term written down explicitly in Eq.\ (\ref{eq:PhysSide}) is a
contribution of the ground-state particle $\mathcal{M}$, whereas
contributions coming from higher resonances and continuum states are denoted
by dots.

The function $\Pi ^{\mathrm{Phys}}(p)$ can be rewritten in a more convenient
form using the matrix element
\begin{equation}
\langle 0|J|\mathcal{M}(p)\rangle =fm.  \label{eq:MElem1}
\end{equation}%
Then, it is not difficult to find $\Pi ^{\mathrm{Phys}}(p)$ in terms of the
parameters $m$ and $f$ \
\begin{equation}
\Pi ^{\mathrm{Phys}}(p)=\frac{m^{2}f^{2}}{m^{2}-p^{2}}+\cdots .
\label{eq:PhysSide1}
\end{equation}%
The Lorentz structure of $\Pi ^{\mathrm{Phys}}(p)$ has a simple form and
consists of a term proportional to $\mathrm{I}$. Then, the invariant
amplitude $\Pi ^{\mathrm{Phys}}(p^{2})$ corresponding to this structure is
given by the expression in the right-hand side of Eq.\ (\ref{eq:PhysSide1}).

The QCD side of the sum rules $\Pi ^{\mathrm{OPE}}(p)$ is determined by Eq.\
(\ref{eq:CF1}) calculated using the $c$ and $s$-quarks propagators. To this
end, we insert the explicit form of $J(x)$ into Eq.\ (\ref{eq:CF1}),
contract heavy and light quark fields, and write the obtained expression
using quark propagators. After these operations, we get
\begin{eqnarray}
&&\Pi ^{\mathrm{OPE}}(p)=i\int d^{4}xe^{ipx}\mathrm{Tr}\left[ \gamma
_{5}S_{c}^{aa^{\prime }}(x)\right.  \notag \\
&&\left. \times \gamma _{5}S_{s}^{a^{\prime }a}(-x)\right] \mathrm{Tr}\left[
\gamma _{5}S_{c}^{b^{\prime }b}(-x)\gamma _{5}S_{s}^{bb^{\prime }}(x)\right]
.  \label{eq:QCDside}
\end{eqnarray}%
Here $S_{c}(x)$ and $S_{s}(x)$ are the $c$ and $s$-quark propagators,
explicit expressions of which can be found in Ref.\ \cite{Agaev:2020zad}.

The correlation function $\Pi ^{\mathrm{OPE}}(p)$ is calculated by employing
the quark propagators with some fixed accuracy of the operator product
expansion ($\mathrm{OPE}$). The $\Pi ^{\mathrm{OPE}}(p)$ has a trivial
Lorentz structure $\sim \mathrm{I}$ as well. Having denoted the
corresponding invariant amplitude by $\Pi ^{\mathrm{OPE}}(p^{2})$ and
equated it to $\Pi ^{\mathrm{Phys}}(p^{2})$, we get a sum rule equality,
which can undergo further processing. The ground-state term and ones due to
higher resonances and continuum states contribute to this sum rule equality
on equal footing. There is a necessity to suppress unwanted contributions of
higher resonances and subtract them from this expression. For these
purposes, we apply the Borel transformation to both its sides. This
operation suppresses effects of higher resonances and continuum states, but
at the same time generates dependence of the obtained equality on the Borel
parameter $M^{2}$. Afterwards, using the assumption about quark-hadron
duality, we perform continuum subtraction, which leads to additional
parameter $s_{0}$ in formulas.

The Borel transformation of the main term in $\Pi ^{\mathrm{Phys}}(p^{2})$
has a simple form
\begin{equation}
\Pi ^{\mathrm{Phys}}(M^{2})=m^{2}f^{2}e^{-m^{2}/M^{2}}.
\end{equation}%
For the Borel transformed and subtracted amplitude $\Pi ^{\mathrm{OPE}%
}(p^{2})$, we find
\begin{equation}
\Pi (M^{2},s_{0})=\int_{4(m_{c}+m_{s})^{2}}^{s_{0}}ds\rho ^{\mathrm{OPE}%
}(s)e^{-s/M^{2}}+\Pi (M^{2}).  \label{eq:InvAmp}
\end{equation}%
The first term in Eq.\ (\ref{eq:InvAmp}) contains an essential part of
contributions, and is expressed using the two-point spectral density $\rho ^{%
\mathrm{OPE}}(s)$, derived as an imaginary part of the correlation function.
The second term $\Pi (M^{2})$ collects nonperturbative contributions
extracted directly from $\Pi ^{\mathrm{OPE}}(p)$.

The sum rules for $m$ and $f$ \ read
\begin{equation}
m^{2}=\frac{\Pi ^{\prime }(M^{2},s_{0})}{\Pi (M^{2},s_{0})}  \label{eq:Mass}
\end{equation}%
and
\begin{equation}
f^{2}=\frac{e^{m^{2}/M^{2}}}{m^{2}}\Pi (M^{2},s_{0}),  \label{eq:Coupling}
\end{equation}%
where $\Pi ^{\prime }(M^{2},s_{0})=d\Pi (M^{2},s_{0})/d(-1/M^{2})$.

Numerical calculations of $m$ and $f$ should be performed in accordance with
Eqs.\ (\ref{eq:Mass}) and (\ref{eq:Coupling}), but only after fixing
different vacuum condensates and working windows for the parameters $M^{2}$
and $s_{0}$. The quark, gluon and mixed condensates are universal and
well-known quantities \cite%
{Shifman:1978bx,Shifman:1978by,Ioffe:1981kw,Ioffe:2005ym,Narison:2015nxh}.
Their numerical values, extracted from numerous processes are listed below
\begin{eqnarray}
&&\langle \overline{q}q\rangle =-(0.24\pm 0.01)^{3}~\mathrm{GeV}^{3},\
\langle \overline{s}s\rangle =(0.8\pm 0.1)\langle \overline{q}q\rangle ,
\notag \\
&&\langle \overline{s}g_{s}\sigma Gs\rangle =m_{0}^{2}\langle \overline{s}%
s\rangle ,\ m_{0}^{2}=(0.8\pm 0.1)~\mathrm{GeV}^{2},\   \notag \\
&&\langle \frac{\alpha _{s}G^{2}}{\pi }\rangle =(0.012\pm 0.004)~\mathrm{GeV}%
^{4},  \notag \\
&&\langle g_{s}^{3}G^{3}\rangle =(0.57\pm 0.29)~\mathrm{GeV}^{6},  \notag \\
&&m_{c}=(1.27\pm 0.02)~\mathrm{GeV},\ \ \ m_{s}=93_{-5}^{+11}~\mathrm{MeV}.
\label{eq:Parameters}
\end{eqnarray}%
We have included the masses of the $c$ and $s$-quarks into Eq.\ (\ref%
{eq:Parameters}) as well. The correlation function $\Pi (M^{2},s_{0})$ is
calculated by taking into account vacuum condensates up to dimension ten.
The expression of $\Pi (M^{2},s_{0})$ is rather lengthy, therefore we do not
provide it here explicitly. In numerical analysis we set $m_{s}^{2}=0$, but
take into account contributions proportional to $m_{s}$.

\begin{figure}[h]
\includegraphics[width=8.5cm]{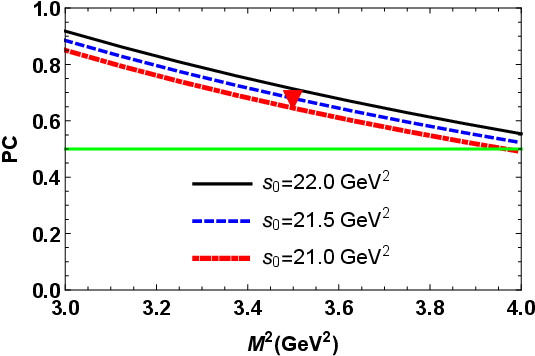}
\caption{Dependence of the pole contribution $\mathrm{PC}$ on the Borel
parameter $M^{2}$ at different $s_{0}$. The limit $\mathrm{PC}=0.5$ is shown
by the horizontal black line. The red triangle shows the point, where the
mass $m$ of the molecule $\mathcal{M}=D_{s}^{+}D_{s}^{-}$ has effectively
been calculated. }
\label{fig:PC}
\end{figure}

To carry out numerical analysis one also needs to choose working regions for
the Borel and continuum subtraction parameters $M^{2}$ and $s_{0}$. They
should satisfy standard constraints of sum rule calculations. Thus, the
parameters $M^{2}$ and $s_{0}$ employed in calculations have to guarantee
the dominance of the pole contribution ($\mathrm{PC}$) and convergence of $%
\mathrm{OPE}$. The former can be defined by the expression
\begin{equation}
\mathrm{PC}=\frac{\Pi (M^{2},s_{0})}{\Pi (M^{2},\infty )},  \label{eqPC}
\end{equation}%
whereas to make sure the operator product expansion converges, we utilize
the ratio
\begin{equation}
R(M^{2})=\frac{\Pi ^{\mathrm{DimN}}(M^{2},s_{0})}{\Pi (M^{2},s_{0})},
\label{eq:Convergence}
\end{equation}%
where $\Pi ^{\mathrm{DimN}}(M^{2},s_{0})$ is a sum of a few last terms in $%
\mathrm{OPE}$.

The $R(M^{2})$ and $\mathrm{PC}$ are used to restrict the lower and upper
bounds for the Borel parameter, respectively. In fact, at $M_{\mathrm{min}%
}^{2}$ the function $R(M_{\mathrm{min}}^{2})$ should be less than some fixed
value, whereas at $M_{\mathrm{max}}^{2}$ the pole contribution $\mathrm{PC}$
has to overshoot the minimally acceptable limit for this parameter. In our
present analysis, we impose on $\mathrm{PC}$ and $R(M^{2})$ the following
constraints
\begin{equation}
\mathrm{PC}\geq 0.5,\ \ R(M_{\mathrm{min}}^{2})\leq 0.05.  \label{eq:Const}
\end{equation}%
The first criterion in Eq.\ (\ref{eq:Const}) is usual for investigations of
conventional hadrons and ensures the dominance of the pole contribution. It
may be used in the analysis of multiquark hadrons as well. But in the case
of multiquark particles this constraint reduces a window for $M^{2}$. The
second condition is required to enforce the convergence of $\mathrm{OPE}$.

Dominance of a perturbative contribution to $\Pi (M^{2},s_{0})$ over a
nonperturbative term, as well as maximum stability of extracted physical
quantities when varying $M^{2}$, is among constraints to determine working
regions for the parameters $M^{2}$ and $s_{0}$. Numerical tests carried out
by taking into account these aspects of the sum rule analysis demonstrate
that the windows for $M^{2}$ and $s_{0}$,
\begin{equation}
M^{2}\in \lbrack 3,4]~\mathrm{GeV}^{2},\ s_{0}\in \lbrack 21,22]~\mathrm{GeV}%
^{2},  \label{eq:Wind1}
\end{equation}%
comply with aforementioned restrictions. Indeed, at $M^{2}=4~\mathrm{GeV}%
^{2} $ the pole contribution on average in $s_{0}$ equals to $0.52$, and
amounts to $0.88$ at $M^{2}=3~\mathrm{GeV}^{2}$. To visualize dynamics of
the pole contribution, in Fig.\ \ref{fig:PC} we depict $\mathrm{PC}$ as a
function of $M^{2}$ at different $s_{0}$. One can see that the pole
contribution overshoots $0.5$ for all values of the parameters $M^{2}$ and $%
s_{0}$ from Eq.\ (\ref{eq:Wind1}).

To be convinced in convergence of $\mathrm{OPE}$, we calculate $\ R(M_{%
\mathrm{min}}^{2})$ at the minimum point $M_{\mathrm{min}}^{2}=3~\mathrm{GeV}%
^{2}$ using in Eq.\ (\ref{eq:Convergence}) dimension-eight, -nine, and -ten
contributions. At the minimal value of $M^{2}$, we find $R(3~\mathrm{GeV}%
^{2})\approx 0.01$ in accordance with the constraint from Eq.\ (\ref%
{eq:Wind1}). Results of a more detailed analysis are shown in Fig.\ \ref%
{fig:Conv}, where one sees contributions to the correlation function $\Pi
(M^{2},s_{0})$ arising from the perturbative and nonperturbative terms up to
dimension eight. The perturbative contribution forms the $0.65$ part of $\Pi
(M^{2},s_{0})$ at $M^{2}=3~\mathrm{GeV}^{2}$ and exceeds the sum of
nonperturbative terms in the whole region of $M^{2}$. The $\mathrm{Dim3}$
term overshoots effects of other nonperturbative operators, which enter to $%
\Pi (M^{2},s_{0})$ with different signs. The $\mathrm{Dim9}$ and $\mathrm{%
Dim10}$ terms are numerically very small and not demonstrated in the figure.

Our predictions for the mass $m$ and coupling $f$ read
\begin{eqnarray}
m &=&(4117\pm 85)~\mathrm{MeV},  \notag \\
f &=&(5.9\pm 0.7)\times 10^{-3}~\mathrm{GeV}^{4}.  \label{eq:Result1}
\end{eqnarray}%
The results for $m$ and $f$ are calculated as their values averaged over the
working regions Eq.\ (\ref{eq:Wind1}). Effectively they correspond to the
sum rule predictions at $M^{2}=3.5~\mathrm{GeV}^{2}$ and $s_{0}=21.5~\mathrm{%
GeV}^{2}$, which is a middle point of the regions of Eq.\ (\ref{eq:Wind1}):
the red triangle in Fig.\ \ref{fig:PC} marks exactly this point. The pole
contribution there is equal to $\mathrm{PC}\approx 0.68$, which in
conjunction with other constraints ensures the ground-state nature of the
molecule $D_{s}^{+}D_{s}^{-}$ and credibility of obtained predictions.

\begin{widetext}

\begin{figure}[h!]
\begin{center}
\includegraphics[totalheight=6cm,width=8cm]{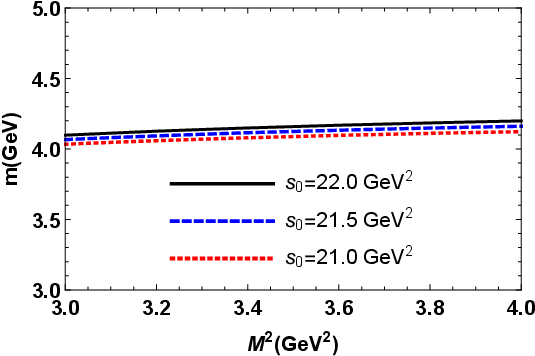}
\includegraphics[totalheight=6cm,width=8cm]{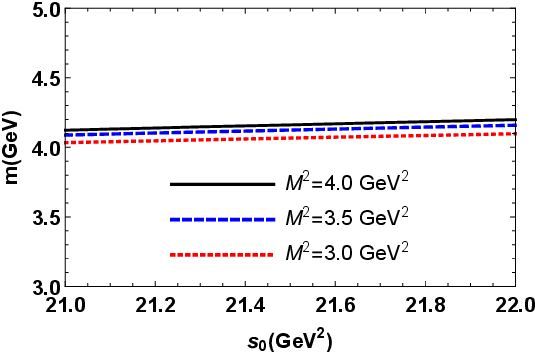}
\end{center}
\caption{Mass $m$ of the molecule $\mathcal{M}$ as a function of the Borel $M^{2}$ (left panel), and continuum threshold $s_0$ parameters (right panel).}
\label{fig:Mass}
\end{figure}

\end{widetext}

In Fig.\ \ref{fig:Mass}, we depict the mass $m$ of the molecule $%
D_{s}^{+}D_{s}^{-}$ as a function of the Borel and continuum subtraction
parameters. Physical quantities obtained from the sum rule analysis should
be stable against variations of the Borel parameter. But the mass $m$
depends on working windows chosen for their calculations. In fact, although
the region for the Borel parameter $M^{2}$ \ leads to an approximately
stable prediction for $m$, there is still residual dependence on it. This
effect generates theoretical uncertainties of the sum rule calculations. It
is worth noting that the uncertainties of $m$ are smaller than ones for the
coupling $f$. The reason is that the mass $m$ is given by the ratio of the
correlation functions which compensates changes of $m$ against $M^{2}$ and $%
s_{0}$. The coupling $f$ , at the same time, depends on $\Pi (M^{2},s_{0})$
and is open for an impact of the parameters $M^{2}$ and $s_{0}$. As a
result, uncertainties of calculations are equal to $\pm 2\%$ in the case of
the mass, and to $\pm 12\%$ for the current coupling.

The region for $s_{0}$ together with $M^{2}$ has to provide the dominance of
$\mathrm{PC}$ and convergence of the operator product expansion. The
parameter $\sqrt{s_{0}}$ also carries useful information on a mass $m^{\ast
} $ of the first radial excitation of the molecule $D_{s}^{+}D_{s}^{-}$.
Thus, in the "ground-state +continuum" scheme adopted in the present work, $%
\sqrt{s_{0}}$ should be less than the mass $m^{\ast }$ of the first excited
state. This fact allows us to estimate the low limit for $m^{\ast }\geq
m+480~\mathrm{MeV}$.

\begin{figure}[h]
\includegraphics[width=8.5cm]{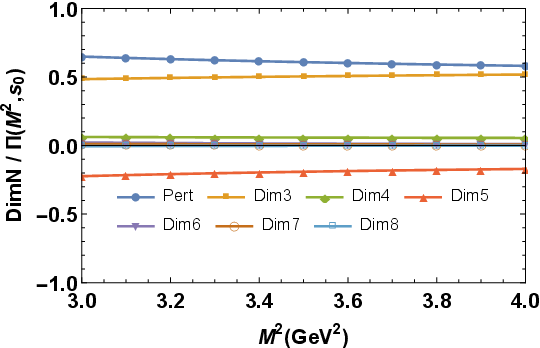}
\caption{Normalized contributions to $\Pi (M^{2},s_{0})$ as functions of the
Borel parameter $M^2$. All curves have been calculated at $s_0=21.5~\mathrm{%
GeV}^2$. }
\label{fig:Conv}
\end{figure}

Our result for the mass of the molecule $\mathcal{M}$ overshoots the LHCb
datum $m_{1\mathrm{exp}}$, but nicely agrees with $m_{2\mathrm{exp}}$.


\section{Partial width of the process $\mathcal{M}\rightarrow
D_{s}^{+}D_{s}^{-}$}

\label{sec:Decays}
The mass and current coupling of the molecule $\mathcal{M}$ calculated in
the previous section provide information to select its possible decay modes.
Besides, one should take into account its quantum numbers $J^{\mathrm{PC}%
}=0^{++}$. Because the structures $X(3960)$ and $X_{0}(4140)$ were
discovered in the invariant mass distribution of the $D_{s}^{+}D_{s}^{-}$
mesons, we consider $\mathcal{M}\rightarrow D_{s}^{+}D_{s}^{-}$ as a
dominant decay mode of $\mathcal{M}$. The two-meson threshold for this decay
is equal approximately to $3937~\mathrm{MeV}$, which makes $\mathcal{M}%
\rightarrow D_{s}^{+}D_{s}^{-}$ the kinematically allowed channel for $%
\mathcal{M}$.

The partial width of the decay $\mathcal{M}\rightarrow D_{s}^{+}D_{s}^{-}$
is governed by a coupling $g_{1}$ that describes strong interaction at the
vertex $\mathcal{M}D_{s}^{+}D_{s}^{-}$. This partial width depends also on
masses and decay constants of the molecule $\mathcal{M}$ and mesons $%
D_{s}^{+}$ and $D_{s}^{-}$. The mass and current coupling of $\mathcal{M}$ \
have been calculated in the present article, whereas physical parameters of
the mesons $D_{s}^{+}$ and $D_{s}^{-}$ are known from independent sources.
Therefore, the only physical quantity to be found is the strong coupling $%
g_{1}$.

To determine $g_{1}$, we employ the QCD three-point sum rule method, and
begin our exploration from the correlation function
\begin{eqnarray}
&&\Pi (p,p^{\prime })=i^{2}\int d^{4}xd^{4}ye^{i(p^{\prime }y-px)}\langle 0|%
\mathcal{T}\{J^{D_{s}^{+}}(y)  \notag \\
&&\times J^{D_{s}^{-}}(0)J^{\dagger }(x)\}|0\rangle .  \label{eq:CF2}
\end{eqnarray}%
Here, $J(x)$,$\ J^{D_{s}^{+}}(y)$ and $J^{D_{s}^{-}}(0)$ are the
interpolating currents for $\mathcal{M}$ and pseudoscalar mesons $D_{s}^{+}$%
\ and $D_{s}^{-}$, respectively. The currents $J^{D_{s}^{+}}$ and $%
J^{D_{s}^{-}}$ are given by the expressions
\begin{eqnarray}
\ J^{D_{s}^{+}}(x) &=&\overline{s}_{j}(x)i\gamma _{5}c_{j}(x),  \notag \\
J^{D_{s}^{-}}(x) &=&\overline{c}_{i}(x)i\gamma _{5}s_{i}(x),  \label{eq:C2}
\end{eqnarray}%
where $i$ and $j$ are color indices. The four-momenta of $\mathcal{M}$ and $%
D_{s}^{+}$ are labeled by $p$ and $p^{\prime }$: Then, the momentum of the
meson $D_{s}^{-}$ is equal to $q=p-p^{\prime }$.

To find $g_{1}$, we apply standard recipes of the sum rule method and
calculate the correlation function $\Pi (p,p^{\prime })$. For these
purposes, we use the physical parameters of the molecule $\mathcal{M}$ and
mesons involved in this decay. The correlator $\Pi (p,p^{\prime })$ obtained
by this manner forms the physical side $\Pi ^{\mathrm{Phys}}(p,p^{\prime })$
of the sum rule. It is easy to see, that
\begin{eqnarray}
&&\Pi ^{\mathrm{Phys}}(p,p^{\prime })=\frac{\langle
0|J^{D_{s}^{+}}|D_{s}^{+}(p^{\prime })\rangle \langle
0|J^{D_{s}^{-}}|D_{s}^{-}(q)\rangle }{(p^{\prime
2}-m_{D_{s}}^{2})(q^{2}-m_{D_{s}}^{2})}  \notag \\
&&\times \frac{\langle D_{s}^{-}(q)D_{s}^{+}(p^{\prime })|\mathcal{M}%
(p)\rangle \langle \mathcal{M}(p)|J^{\dagger }|0\rangle }{(p^{2}-m^{2})}%
+\cdots ,  \notag \\
&&  \label{eq:CF3}
\end{eqnarray}%
with $m_{D_{s}}$ being the mass of the mesons $D_{s}^{\pm }$. To get Eq.\ (%
\ref{eq:CF3}), we isolate contributions of the ground-state and higher
resonances and continuum state particles from each other. In Eq.\ (\ref%
{eq:CF3}) the ground-state term is written down explicitly, whereas other
contributions are denoted by ellipses.

The function $\Pi ^{\mathrm{Phys}}(p,p^{\prime })$ can be rewritten in terms
of the $D_{s}^{\pm }$ mesons matrix elements
\begin{equation}
\langle 0|J^{D_{s}^{\pm }}|D_{s}^{\pm }\rangle =\frac{m_{D_{s}}^{2}f_{D_{s}}%
}{m_{c}+m_{s}},\   \label{eq:Mel2}
\end{equation}%
where $f_{D_{s}}$ is their decay constant. We model the vertex $\mathcal{M}%
D_{s}^{+}D_{s}^{-}$ by the matrix element%
\begin{equation}
\langle D_{s}^{-}(q)D_{s}^{+}(p^{\prime })|\mathcal{M}(p)\rangle
=g_{1}(q^{2})p\cdot p^{\prime }.  \label{eq:Ver1}
\end{equation}%
Using these expressions, it is not difficult to calculate the new expression
of the correlation function $\Pi ^{\mathrm{Phys}}(p,p^{\prime })$
\begin{eqnarray}
&&\Pi ^{\mathrm{Phys}}(p,p^{\prime })=g_{1}(q^{2})\frac{%
m_{D_{s}}^{4}f_{D_{s}}^{2}fm}{(m_{c}+m_{s})^{2}(p^{2}-m^{2})}  \notag \\
&&\times \frac{1}{(p^{\prime 2}-m_{D_{s}}^{2})(q^{2}-m_{D_{s}}^{2})}\frac{%
m^{2}+m_{D_{s}}^{2}-q^{2}}{2}+\cdots .  \notag \\
&&  \label{eq:Phys2}
\end{eqnarray}%
The double Borel transformation of the function $\Pi ^{\mathrm{Phys}%
}(p,p^{\prime })$ over variables $p^{2}$ and $p^{\prime 2}$ is given by the
formula
\begin{eqnarray}
&&\mathcal{B}\Pi ^{\mathrm{Phys}}(p,p^{\prime })=g_{1}(q^{2})\frac{%
m_{D_{s}}^{4}f_{D_{s}}^{2}fm}{(m_{c}+m_{s})^{2}(q^{2}-m_{D_{s}}^{2})}%
e^{-m^{2}/M_{1}^{2}}  \notag \\
&&\times e^{-m_{D_{s}}^{2}/M_{2}^{2}}\frac{m^{2}+m_{D_{s}}^{2}-q^{2}}{2}%
+\cdots .  \label{eq:BTr}
\end{eqnarray}%
The correlator $\Pi ^{\mathrm{Phys}}(p,p^{\prime })$ and its Borel
transformation has simple Lorentz structure $\sim \mathrm{I}$. Then the
whole expression in Eq.\ (\ref{eq:Phys2}) determines the invariant amplitude
$\Pi ^{\mathrm{Phys}}(p^{2},p^{\prime 2},q^{2})$.

To find the QCD side of the three-point sum rule, we express $\Pi
(p,p^{\prime })$ in terms of quark propagators, and get
\begin{eqnarray}
&&\Pi ^{\mathrm{OPE}}(p,p^{\prime })=i^{2}\int d^{4}xd^{4}ye^{i(p^{\prime
}y-px)}  \notag \\
&&\times \mathrm{Tr}\left[ \gamma _{5}S_{c}^{ia}(y-x)\gamma
_{5}S_{s}^{ai}(x-y)\right]  \notag \\
&&\times \mathrm{Tr}\left[ \gamma _{5}S_{s}^{jb}(-x)\gamma _{5}S_{c}^{bj}(x)%
\right].  \label{eq:CF4}
\end{eqnarray}

The correlator $\Pi ^{\mathrm{OPE}}(p,p^{\prime })$ is computed by taking
into account nonperturbative contributions up to dimension $6$, and as $\Pi
^{\mathrm{Phys}}(p,p^{\prime })$, contains the Lorentz structure
proportional to \textrm{I}. Denoting relevant invariant amplitude by $\Pi ^{%
\mathrm{OPE}}(p^{2},p^{\prime 2},q^{2})$ , equating its double Borel
transformation $\mathcal{B}\Pi ^{\mathrm{OPE}}(p^{2},p^{\prime 2},q^{2})$ to
$\mathcal{B}\Pi ^{\mathrm{Phys}}(p^{2},p^{\prime 2},q^{2})$, and performing
continuum subtraction, we get the sum rule for the coupling $g_{1}(q^{2})$.

After these manipulations, the amplitude $\Pi ^{\mathrm{OPE}%
}(p^{2},p^{\prime 2},q^{2})$ can be rewritten using the spectral density $%
\rho (s,s^{\prime },q^{2})$, which is extracted as an imaginary part of $\Pi
^{\mathrm{OPE}}(p,p^{\prime })$
\begin{eqnarray}
&&\Pi (\mathbf{M}^{2},\mathbf{s}_{0},q^{2})=%
\int_{4(m_{c}+m_{s})^{2}}^{s_{0}}ds\int_{(m_{c}+m_{s})^{2}}^{s_{0}^{\prime
}}ds^{\prime }\rho (s,s^{\prime },q^{2})  \notag \\
&&\times e^{-s/M_{1}^{2}}e^{-s^{\prime }/M_{2}^{2}},  \label{eq:SCoupl}
\end{eqnarray}%
where $\mathbf{M}^{2}=(M_{1}^{2},\ M_{2}^{2})$ and $\mathbf{s}_{0}=(s_{0},\
s_{0}^{\prime })$ are the Borel and continuum threshold parameters. The
couples $(M_{1}^{2},\ s_{0})$ and $\mathbf{s}_{0}=(M_{2}^{2},\ s_{0}^{\prime
})$ correspond to the molecule $M$ and $D_{s}^{+}$ meson channels,
respectively.

The sum rule for $g_{1}(q^{2})$ is determined by the formula
\begin{eqnarray}
g_{1}(q^{2}) &=&\frac{2(m_{c}+m_{s})^{2}}{m_{D_{s}}^{4}f_{D_{s}}^{2}fm}\frac{%
q^{2}-m_{D_{s}}^{2}}{m^{2}+m_{D_{s}}^{2}-q^{2}}  \notag \\
&&\times e^{m^{2}/M_{1}^{2}}e^{m_{D_{s}}^{2}/M_{2}^{2}}\Pi (\mathbf{M}^{2},%
\mathbf{s}_{0},q^{2}).  \label{eq:SRCoup}
\end{eqnarray}

The expression of $g_{1}(q^{2})$ depends on the spectroscopic parameters of
the molecule $\mathcal{M}$, as well as the masses and decay constants of the
mesons $D_{s}^{\pm }$: they are input parameters of numerical computations.
Values of these parameters as well as masses and decay constants $f_{D_{s}}$%
, $f_{\eta _{c}}$, $f_{\phi }$ and $f_{J/\psi }$ that are necessary to study
other decays are collected in Table\ \ref{tab:Param}. The masses all of
mesons are borrowed from Ref.\ \cite{PDG:2022}. For the decay constant of
the mesons $D_{s}^{\pm }$, we employ information from the same source,
whereas for $f_{\eta _{c}}$ use a prediction made in Ref.\ \cite%
{Colangelo:1992cx} on the basis of the sum rule method. As the decay
constants $f_{\phi }$ and $f_{J/\psi }$ of the vector mesons $\phi $ and $%
J/\psi $, we utilize the experimental values reported in Refs. \cite%
{Chakraborty:2017hry,Kiselev:2001xa}, respectively.

The partial width of the decay $\mathcal{M}\rightarrow D_{s}^{+}D_{s}^{-}$
besides various input parameters also is determined by the strong coupling $%
g_{1}(m_{D_{s}}^{2})$ at the mass shell $q^{2}=m_{D_{s}}^{2}$ of the meson $%
D_{s}^{-}$. At the same time, the sum rule computations of $g_{1}$ can be
carried out in deep-Euclidean region $q^{2}<0$. For simplicity, we introduce
a variable $Q^{2}=-q^{2}$ and in what follows label the obtained function by
$g_{1}(Q^{2})$. An explored range of $Q^{2}$ covers the region $Q^{2}=1-5~%
\mathrm{GeV}^{2}$.

Numerical calculations also require choosing the working regions for the
Borel and continuum subtraction parameters $\mathbf{M}^{2}$ and $\mathbf{s}%
_{0}$. Limits imposed on $\mathbf{M}^{2}$ and $\mathbf{s}_{0}$ are standard
for sum rule calculations and were considered in the previous section. The
regions for $M_{1}^{2}$ and $s_{0}$ associated with the $\mathcal{M}$
channel are fixed in accordance with Eq.\ (\ref{eq:Wind1}). The parameters $%
(M_{2}^{2},\ s_{0}^{\prime })$ for the $D_{s}^{+}$ meson channel are varied
within limits
\begin{equation}
M_{2}^{2}\in \lbrack 2.5,3.5]~\mathrm{GeV}^{2},\ s_{0}^{\prime }\in \lbrack
5,6]~\mathrm{GeV}^{2}.  \label{eq:Wind3}
\end{equation}%
Regions for $\mathbf{M}^{2}$ and $\mathbf{s}_{0}$ are chosen in such a way
to minimize their effects of the coupling $g_{1}(Q^{2})$.

\begin{figure}[h]
\includegraphics[width=8.5cm]{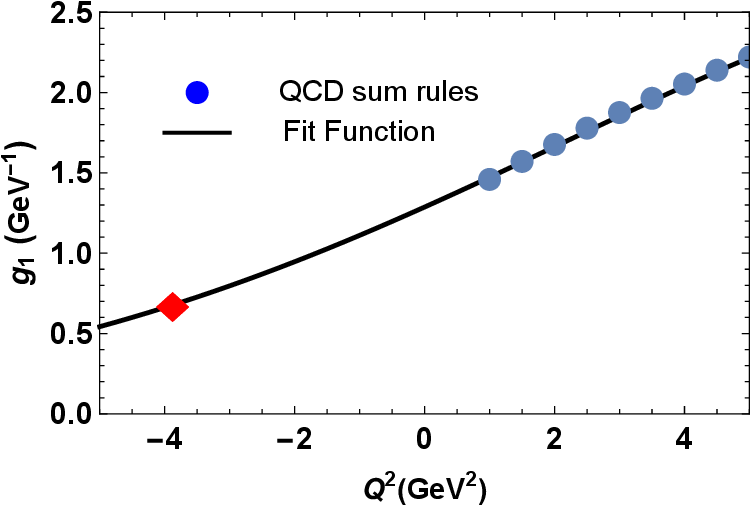}
\caption{The sum rule results and fit function for the strong coupling $%
g_{1}(Q^{2})$. The red diamond denotes the point $Q^{2}=-m_{D_s}^{2}$. }
\label{fig:Fit}
\end{figure}

Results of computations are pictured in Fig.\ \ref{fig:Fit}. It is seen that
results for $g_{1}(Q^{2})$ are extracted at the region $Q^{2}>0$, where the
sum rule gives reliable predictions. It has just been explained above that
we need $g_{1}(Q^{2})$ at $Q^{2}=-m_{D_{s}}^{2}$. To this end, one has to
introduce some fit function $\mathcal{G}_{1}(Q^{2})$, which at the momenta $%
Q^{2}>0$ gives the same values as the sum rule computations, but can easily
be extrapolated to the region $Q^{2}<0$. There are different choices for
such functions. In this article, we use $\mathcal{G}_{i}(Q^{2}),\ i=1,2,3$
\begin{equation}
\mathcal{G}_{i}(Q^{2})=\mathcal{G}_{i}^{0}\mathrm{\exp }\left[ c_{i}^{1}%
\frac{Q^{2}}{m^{2}}+c_{i}^{2}\left( \frac{Q^{2}}{m^{2}}\right) ^{2}\right] ,
\label{eq:FitF}
\end{equation}%
where $\mathcal{G}_{i}^{0}$, $c_{i}^{1}$, and $c_{i}^{2}$ are parameters,
which should be extracted from fitting procedures. Calculations demonstrate
that $\mathcal{G}_{1}^{0}=1.29~\mathrm{GeV}^{-1}$, $c_{1}^{1}=2.38$, and $%
c_{1}^{2}=-1.84$ lead to a reasonable agreement with the sum rule's data
(see, Fig.\ \ref{fig:Fit}).

At the mass shell $q^{2}=m_{D_{s}}^{2}$ this function gives
\begin{equation}
g_{1}\equiv \mathcal{G}_{1}(-m_{D_{s}}^{2})=(6.8\pm 1.6)\times 10^{-1}\
\mathrm{GeV}^{-1}.  \label{eq:Coupl1}
\end{equation}%
The width of the decay $\mathcal{M}\rightarrow D_{s}^{+}D_{s}^{-}$ is
calculated by means of the formula%
\begin{equation}
\Gamma \left[ \mathcal{M}\rightarrow D_{s}^{+}D_{s}^{-}\right] =g_{1}^{2}%
\frac{m_{D_{s}}^{2}\lambda }{8\pi }\left( 1+\frac{\lambda ^{2}}{m_{D_{s}}^{2}%
}\right) ,  \label{eq:PartDW}
\end{equation}%
where $\lambda =\lambda \left( m,m_{D_{s}},m_{D_{s}}\right) $ and
\begin{eqnarray}
\lambda \left( a,b,c\right) &=&\frac{1}{2a}\left[ a^{4}+b^{4}+c^{4}\right.
\notag \\
&&\left. -2\left( a^{2}b^{2}+a^{2}c^{2}+b^{2}c^{2}\right) \right] ^{1/2}.
\end{eqnarray}%
Using the coupling Eq.\ (\ref{eq:Coupl1}), it is not difficult to compute
the width of the decay $\mathcal{M}\rightarrow D_{s}^{+}D_{s}^{-}$%
\begin{equation}
\Gamma \left[ \mathcal{M}\rightarrow D_{s}^{+}D_{s}^{-}\right] =(46.5\pm
11.6)~\mathrm{MeV}.  \label{eq:DW1Numeric}
\end{equation}

\begin{table}[tbp]
\begin{tabular}{|c|c|}
\hline\hline
Quantity & Value (in $\mathrm{MeV}$ units) \\ \hline
$m_{D_s}$ & $1969.0\pm 1.4$ \\
$m_{\eta_{c}}$ & $2983.9 \pm 0.4$ \\
$m_{\eta^{\prime}}$ & $957.78 \pm 0.06$ \\
$m_{\eta}$ & $547.862 \pm 0.017$ \\
$m_{J/\psi}$ & $3096.900 \pm 0.006$ \\
$m_{\phi}$ & $1019.461 \pm 0.019$ \\
$f_{D_s}$ & $249.9 \pm 0.5$ \\
$f_{\eta_c}$ & $320 \pm 40$ \\
$f_{J/\psi}$ & $409 \pm 15$ \\
$f_{\phi}$ & $228.5 \pm 3.6 $ \\ \hline\hline
\end{tabular}%
\caption{Masses and decay constants of the mesons $D_{s}^{\pm}$, $\protect%
\eta_c$, $\protect\eta^{\prime}$, $\protect\eta$, $J/\protect\psi$ and $%
\protect\phi$ which are employed in numerical calculations.}
\label{tab:Param}
\end{table}

\section{Decays $\mathcal{M}\rightarrow \protect\eta _{c}\protect\eta %
^{\prime }$, $\ \mathcal{M}\rightarrow \protect\eta _{c}\protect\eta $ \ and
$\mathcal{M}\rightarrow J/\protect\psi \protect\phi $}

\label{sec:DecaysA}

Other processes that we study here are decays $\mathcal{M}\rightarrow \eta
_{c}\eta ^{\prime }$, $\mathcal{M}\rightarrow \eta _{c}\eta $, and $\mathcal{%
M}\rightarrow J/\psi \phi $. The two-meson threshold $3941~\mathrm{MeV}$ for
the first two decays is below the mass $m$ of the molecule $\mathcal{M}$.
The threshold for the decay $\mathcal{M}\rightarrow J/\psi \phi $ is less
than $m$ as well. It is easy to prove that these processes conserve the $%
\mathrm{P}$ and $\mathrm{C}$ parities of the initial particle $\mathcal{M}$.


\subsection{$\mathcal{M}\rightarrow \protect\eta _{c}\protect\eta ^{\prime }$
and $\mathcal{M}\rightarrow \protect\eta _{c}\protect\eta $}


The decays $\mathcal{M}\rightarrow \eta _{c}\eta ^{\prime }$ and $\mathcal{M}%
\rightarrow \eta _{c}\eta $ are studied by a method described above. But,
here we take into account peculiarities of the $\eta -\eta ^{\prime }$
mesons connected with mixing in this system due to the $U(1)$ anomaly \cite%
{Feldmann:1998vh}. This effect modifies a choice of interpolating currents
and matrix elements for these particles. Although $\eta -\eta ^{\prime }$
mixing can be considered in different approaches, we use the quark-flavor
basis $|\eta _{q}\rangle =(\overline{u}u+\overline{d}d)/\sqrt{2}$ and $|\eta
_{s}\rangle =\overline{s}s$, where the physical particles\ $\eta $ and $\eta
^{\prime }$ have simple decompositions \cite%
{Feldmann:1998vh,Agaev:2014wna,Agaev:2015faa}%
\begin{eqnarray}
\eta &=&|\eta _{q}\rangle \cos \varphi -|\eta _{s}\rangle \sin \varphi ,
\notag \\
\eta ^{\prime } &=&|\eta _{q}\rangle \sin \varphi +|\eta _{s}\rangle \cos
\varphi ,  \label{eq:PhysMes}
\end{eqnarray}%
where $\varphi $ is the mixing angle in the $\{|\eta _{q}\rangle ,\ |\eta
_{s}\rangle \}$ basis. Such state mixing implies that the same assumption is
applicable to their currents, decay constants and matrix elements.

Because in the decays $\mathcal{M}\rightarrow \eta _{c}\eta ^{\prime }$, $%
\eta _{c}\eta $ participate only $\overline{s}s$ components of the mesons $%
\eta $ and $\eta ^{\prime }$, relevant interpolating currents are given by
the expressions%
\begin{eqnarray}
J^{\eta }(x) &=&-\sin \varphi \overline{s}_{j}(x)i\gamma _{5}s_{j}(x),
\notag \\
J^{\eta ^{\prime }}(x) &=&\cos \varphi \overline{s}_{j}(x)i\gamma
_{5}s_{j}(x),  \label{eq:EtaCurr}
\end{eqnarray}%
where $j$ is the color index.

We start our analysis from the process $\mathcal{M}\rightarrow \eta _{c}\eta
^{\prime }$. Then, we should consider the correlation function
\begin{eqnarray}
\widetilde{\Pi }(p,p^{\prime }) &=&i^{2}\int d^{4}xd^{4}ye^{i(p^{\prime
}y-px)}\langle 0|\mathcal{T}\{J^{\eta _{c}}(y)  \notag \\
&&\times J^{\eta ^{\prime }}(0)J^{\dagger }(x)\}|0\rangle ,  \label{eq:CF2a}
\end{eqnarray}%
where $\ J^{\eta _{c}}(y)$ is the interpolating current of the meson $\eta
_{c}$
\begin{equation}
\ J^{\eta _{c}}(x)=\overline{c}_{i}(x)i\gamma _{5}c_{i}(x).
\label{eq:Curr5a}
\end{equation}%
The main contribution to the correlation function $\widetilde{\Pi }%
(p,p^{\prime })$ has the following form
\begin{eqnarray}
&&\widetilde{\Pi }^{\mathrm{Phys}}(p,p^{\prime })=\frac{\langle 0|J^{\eta
_{c}}|\eta _{c}(p^{\prime })\rangle \langle 0|J^{\eta ^{\prime }}|\eta
^{\prime }(q)\rangle }{(p^{\prime 2}-m_{\eta _{c}}^{2})(q^{2}-m_{\eta
^{\prime }}^{2})}  \notag \\
&&\times \frac{\langle \eta ^{\prime }(q)\eta _{c}(p^{\prime })|\mathcal{M}%
(p)\rangle \langle \mathcal{M}(p)|J^{\dagger }|0\rangle }{p^{2}-m^{2}}%
+\cdots ,  \notag \\
&&  \label{eq:CF5}
\end{eqnarray}%
where the ellipses stand for contributions of higher resonances and
continuum states. The function $\widetilde{\Pi }^{\mathrm{Phys}}(p,p^{\prime
})$ can be rewritten using the matrix elements
\begin{eqnarray}
&&\langle 0|J^{\eta _{c}}|\eta _{c}\rangle =\frac{m_{\eta _{c}}^{2}\ f_{\eta
_{c}}}{2m_{c}},\   \notag \\
&&2m_{s}\langle \eta ^{\prime }|\overline{s}i\gamma _{5}s|0\rangle =h_{\eta
^{\prime }}^{s},  \label{eq:Mel3}
\end{eqnarray}%
where $m_{\eta _{c}}$ and $\ f_{\eta _{c}}$ are the mass and decay constant
of the $\eta _{c}$ meson. The matrix element of local operator $\overline{s}%
i\gamma _{5}s$ sandwiched between the meson $\eta ^{\prime }$ and vacuum
states is denoted by $h_{\eta ^{\prime }}^{s}$ \cite{Agaev:2014wna}. The
parameter $h_{\eta ^{\prime }}^{s}$ follows the $\eta -\eta ^{\prime }$
state-mixing pattern, and we get%
\begin{equation}
h_{\eta ^{\prime }}^{s}=h_{s}\cos \varphi .  \label{eq:Mel4}
\end{equation}%
The parameter $h_{s}$ can be defined theoretically \cite{Agaev:2014wna}, but
for our purposes it is enough to use values of $h_{s}$ and $\varphi $
extracted from phenomenological analyses
\begin{eqnarray}
h_{s} &=&(0.087\pm 0.006)~\mathrm{GeV}^{3},  \notag \\
\varphi &=&39.3^{\circ }\pm 1.0^{\circ }.  \label{eq:MixParam}
\end{eqnarray}

The vertex $\mathcal{M}\eta _{c}\eta ^{\prime }$ has the following form%
\begin{equation}
\langle \eta ^{\prime }(q)\eta _{c}(p^{\prime })|\mathcal{M}(p)\rangle
=g_{2}(q^{2})p\cdot p^{\prime },
\end{equation}%
where $g_{2}$ is the strong coupling at the vertex $\mathcal{M}\eta _{c}\eta
^{\prime }$. By employing these matrix elements, we obtain a new expression
for $\widetilde{\Pi }^{\mathrm{Phys}}(p,p^{\prime })$:
\begin{eqnarray}
&&\widetilde{\Pi }^{\mathrm{Phys}}(p,p^{\prime })=g_{2}(q^{2})\frac{%
fmm_{\eta _{c}}^{2}\ f_{\eta _{c}}h_{s}\cos ^{2}\varphi }{%
4m_{c}m_{s}{}(p^{2}-m^{2})}  \notag \\
&&\times \frac{1}{(p^{\prime 2}-m_{\eta _{c}}^{2})(q^{2}-m_{\eta ^{\prime
}}^{2})}\frac{m^{2}+m_{\eta _{c}}^{2}-q^{2}}{2}+\cdots .  \notag \\
&&
\end{eqnarray}%
The QCD side of the sum rule for $g_{2}(q^{2})$ is given by the formula
\begin{eqnarray}
&&\widetilde{\Pi }^{\mathrm{OPE}}(p,p^{\prime })=-\cos \varphi \int
d^{4}xd^{4}ye^{i(p^{\prime }y-px)}  \notag \\
&&\times \mathrm{Tr}\left[ \gamma _{5}S_{c}^{ia}(y-x)\gamma
_{5}S_{s}^{ai}(x-y)\right]  \notag \\
&&\times \mathrm{Tr}\left[ \gamma _{5}S_{s}^{jb}(-x)\gamma
_{5}S_{c}^{bj}(x-y)\right] .  \label{eq:CF6}
\end{eqnarray}%
The sum rule for the coupling $g_{2}(q^{2})$ is obtained using the Borel
transformations of invariant amplitudes $\widetilde{\Pi }^{\mathrm{Phys}%
}(p^{2},p^{\prime 2},q^{2})$ and $\widetilde{\Pi }^{\mathrm{OPE}%
}(p^{2},p^{\prime 2},q^{2})$ and is equal to%
\begin{eqnarray}
g_{2}(q^{2}) &=&-\frac{8m_{c}m_{s}}{fmm_{\eta _{c}}^{2}\ f_{\eta
_{c}}h_{s}\cos \varphi }\frac{q^{2}-m_{\eta _{c}}^{2}}{m^{2}+m_{\eta
_{c}}^{2}-q^{2}}  \notag \\
&&\times e^{m^{2}/M_{1}^{2}}e^{m_{\eta _{c}}^{2}/M_{2}^{2}}\widetilde{\Pi }(%
\mathbf{M}^{2},\mathbf{s}_{0},q^{2}).  \label{eq:StCoup1}
\end{eqnarray}%
Here, $\widetilde{\Pi }(\mathbf{M}^{2},\mathbf{s}_{0},q^{2})$ is the Borel
transformed and subtracted amplitude $\widetilde{\Pi }^{\mathrm{OPE}%
}(p^{2},p^{\prime 2},q^{2})$.

The coupling $g_{1}(q^{2})$ is calculated using the following Borel and
continuum threshold parameters in the $\eta _{c}$ channel
\begin{equation}
M_{2}^{2}\in \lbrack 3,4]~\mathrm{GeV}^{2},\ s_{0}^{\prime }\in \lbrack
9.5,10.5]~\mathrm{GeV}^{2},  \label{eq.Wind4}
\end{equation}%
whereas for the $\mathcal{M}$ channel, we employ $M_{1}^{2}$ and $s_{0}$ \
from Eq.\ (\ref{eq:Wind1}). The strong coupling $g_{2}$ is defined at the
mass shell of the $\eta ^{\prime }$ meson. The fit function $\mathcal{G}%
_{2}(Q^{2})$ given by Eq.\ (\ref{eq:FitF}) has the parameters $\mathcal{G}%
_{2}^{0}=0.21~\mathrm{GeV}^{-1}$, $c_{2}^{1}=5.08$, and $c_{2}^{2}=-4.04$.
Computations yield%
\begin{equation}
g_{2}\equiv \mathcal{G}_{2}(-m_{\eta ^{\prime }}^{2})=(1.6\pm 0.3)\times
10^{-1}\ \mathrm{GeV}^{-1}.  \label{eq:g1}
\end{equation}

The partial width of this decay can be evaluated using the formula Eq.\ (\ref%
{eq:PartDW}), in which one should make substitutions $g_{1}\rightarrow g_{2}$%
, $m_{D_{s}}^{2}\rightarrow m_{\eta _{c}}^{2}$ and $\lambda \left(
m,m_{D_{s}},m_{D_{s}}\right) \rightarrow \widetilde{\lambda }\left(
m,m_{\eta _{c}},m_{\eta ^{\prime }}\right) $. Then, for the width of the
decay$\mathcal{M}\rightarrow \eta _{c}\eta ^{\prime }$, we find
\begin{equation}
\Gamma \left[ \mathcal{M}\rightarrow \eta _{c}\eta ^{\prime }\right]
=(4.9\pm 1.1)~\mathrm{MeV}.  \label{eq:DW2Numeric}
\end{equation}

Analysis of the decay $\mathcal{M}\rightarrow \eta _{c}\eta $ can be
performed in a similar manner. Avoiding further details, let us write down
the predictions obtained for key quantities. Thus, the strong coupling $%
g_{3} $ at the vertex $\mathcal{M}\eta _{c}\eta $ is given by the formula
\begin{equation}
g_{3}\equiv |\mathcal{G}_{3}(-m_{\eta }^{2})|=(1.5\pm 0.3)\times 10^{-1}\
\mathrm{GeV}^{-1},  \label{eq:g2}
\end{equation}%
where parameters of the fit function are $\mathcal{G}_{3}^{0}=-0.17~\mathrm{%
GeV}^{-1}$, $c_{3}^{1}=7.29$, and $c_{3}^{2}=-7.33$. The width of the decay $%
\mathcal{M}\rightarrow \eta _{c}\eta $ is%
\begin{equation}
\Gamma \left[ \mathcal{M}\rightarrow \eta _{c}\eta \right] =(7.7\pm 1.8)~%
\mathrm{MeV}.  \label{eq:DW3}
\end{equation}


\subsection{$\mathcal{M}\rightarrow J/\protect\psi \protect\phi $}


The process $\mathcal{M}\rightarrow J/\psi \phi $ is the kinematically
allowed decay channel of the molecule $\mathcal{M}$. The hadronic molecule $%
\mathcal{M}$ can decay also to $J/\psi \omega $ mesons, because through a
mixing phenomenon $\omega $ acquires a strange-quark component. As in the
case of $\eta $ and $\eta ^{\prime }$ mesons, the $\omega -\phi $ mixing can
be defined in the following form
\begin{eqnarray}
\omega &=&|\overline{q}q\rangle \cos \psi _{\mathrm{V}}-|\overline{s}%
s\rangle \sin \psi _{\mathrm{V}},  \notag \\
\phi &=&|\overline{q}q\rangle \sin \psi _{\mathrm{V}}+|\overline{s}s\rangle
\cos \psi _{\mathrm{V}},
\end{eqnarray}%
where $|\overline{q}q\rangle $ and $|\overline{s}s\rangle $ are vector
counterparts of the basic states $|\eta _{q}\rangle ,\ |\eta _{s}\rangle $,
and $\psi _{\mathrm{V}}$ is the $\omega -\phi $ mixing angle. But in
contrast to $\varphi $, the mixing angle $\psi _{\mathrm{V}}$ is numerically
very small \cite{Ambrosino:2009sc}
\begin{equation}
\psi _{\mathrm{V}}=(3.32\pm 0.09)^{\circ }.
\end{equation}%
As a result, a $|\overline{s}s\rangle $ component of the meson $\omega $ is
small as well. In other words, the $\phi $ and $\omega $ mesons are almost
purely strange and nonstrange vector particles, respectively. Therefore, we
consider only the decay $\mathcal{M}\rightarrow J/\psi \phi $ and neglect
the contribution to the full width of $\mathcal{M}$ coming from the process $%
\mathcal{M}\rightarrow J/\psi \omega $.

The correlation function to be examined in the case of decay $\mathcal{M}%
\rightarrow J/\psi \phi $ is
\begin{eqnarray}
&&\widehat{\Pi }_{\mu \nu }(p,p^{\prime })=i^{2}\int
d^{4}xd^{4}ye^{i(p^{\prime }y-px)}\langle 0|\mathcal{T}\{J_{\mu }^{J/\psi
}(y)  \notag \\
&&\times J_{\nu }^{\phi }(0)J^{\dagger }(x)\}|0\rangle ,  \label{eq:CF7}
\end{eqnarray}%
where $J_{\mu }^{J/\psi }$ and $J_{\nu }^{\phi }$ are interpolating currents
for the vector mesons $J/\psi $ and $\phi $, respectively. As is seen, the
momenta of the molecule $\mathcal{M}$ and meson $J/\psi $ are equal to $p$
and $p^{\prime }$, respectively. Consequently, the momentum of the $\phi $
meson is $q=p-p^{\prime }$.

The interpolating currents $J_{\mu }^{J/\psi }$ and $J_{\nu }^{\phi }$ are
defined by the following expressions%
\begin{eqnarray}
J_{\mu }^{J/\psi }(x) &=&\overline{c}_{i}(x)\gamma _{\mu }c_{i}(x),  \notag
\\
J_{\nu }^{\phi }(x) &=&\overline{s}_{j}(x)\gamma _{\nu }s_{j}(x).
\label{eq:Curr6}
\end{eqnarray}%
To find the physical side of the sum rule, we express $\widehat{\Pi }_{\mu
\nu }(p,p^{\prime })$ using parameters of the involved particles through
their matrix elements
\begin{eqnarray}
&&\langle 0|J_{\mu }^{J/\psi }|J/\psi (p^{\prime })\rangle =f_{J/\psi
}m_{J/\psi }\varepsilon _{\mu }(p^{\prime }),\   \notag \\
&&\langle 0|J_{\nu }^{\phi }|\phi (q)\rangle =f_{\phi }m_{\phi }\varepsilon
_{\nu }(q).  \label{eq:MEl1}
\end{eqnarray}%
For the vertex $\langle \phi ^{\prime }(q)J/\psi (p^{\prime })|\mathcal{M}%
(p)\rangle $, we employ the matrix element%
\begin{eqnarray}
&&\langle \phi ^{\prime }(q)J/\psi (p^{\prime })|\mathcal{M}(p)\rangle
=g_{4}(q^{2})\left[ p^{\prime }\cdot \varepsilon ^{\ast }(q)\right.  \notag
\\
&&\left. \times q\cdot \varepsilon ^{\ast }(p^{\prime })-p^{\prime }\cdot
q\varepsilon ^{\ast }(q)\cdot \varepsilon ^{\ast }(p^{\prime })\right] .
\label{eq:Vert}
\end{eqnarray}%
In the formulas above, $m_{J/\psi }$, $m_{\phi }$, $\ \ f_{J/\psi }$, \ $%
f_{\phi }$ and $\varepsilon _{\mu }(p^{\prime })$, $\varepsilon _{\nu }(q)$
are the masses, decay constants and polarization vectors of the mesons $%
J_{\mu }^{J/\psi }$ and $J_{\nu }^{\phi }$, respectively. The strong
coupling $g_{4}$ in Eq.\ (\ref{eq:Vert}) corresponds to the vertex $\mathcal{%
M}J/\psi \phi $.

Then the physical side $\widehat{\Pi }_{\mu \nu }^{\mathrm{Phys}%
}(p,p^{\prime })$ of the sum rule takes the form%
\begin{eqnarray}
&&\widehat{\Pi }_{\mu \nu }^{\mathrm{Phys}}(p,p^{\prime })=-g_{4}(q^{2})%
\frac{mm_{J/\psi }m_{\phi }ff_{J/\psi }f_{\phi }}{(p^{2}-m^{2})(p^{\prime
2}-m_{J/\psi }^{2})}  \notag \\
&&\times \frac{1}{(q^{2}-m_{\phi }^{2})}\left[ \frac{m^{2}-m_{J/\psi
}^{2}-q^{2}}{2}g_{\mu \nu }-p_{\nu }^{\prime }q_{\mu }\right] +\cdots .
\notag \\
&&  \label{eq:CF8}
\end{eqnarray}%
The same correlation function in terms of quark propagators is given by the
formula%
\begin{eqnarray}
&&\widehat{\Pi }_{\mu \nu }^{\mathrm{OPE}}(p,p^{\prime })=i^{2}\int
d^{4}xd^{4}ye^{i(p^{\prime }y-px)}  \notag \\
&&\times \mathrm{Tr}\left[ \gamma _{\mu }S_{c}^{ia}(y-x)\gamma
_{5}S_{s}^{aj}(x)\gamma _{\nu }\right.  \notag \\
&&\left. \times S_{s}^{jb}(-x)\gamma _{5}S_{c}^{bi}(x-y)\right] .
\label{eq:CF9}
\end{eqnarray}%
Remaining operations with functions $\widehat{\Pi }_{\mu \nu }^{\mathrm{Phys}%
}(p,p^{\prime })$ and $\widehat{\Pi }_{\mu \nu }^{\mathrm{OPE}}(p,p^{\prime
})$ are standard manipulations of the sum rule analysis. Let us note only
that the sum rule for $g_{4}$ is derived using invariant amplitudes which
correspond to structures $\sim g_{\mu \nu }$ in these correlators. In
numerical analysis the second pair of the parameters $(M_{2}^{2},\
s_{0}^{\prime })$ related to the $J/\psi $ channel is chosen as%
\begin{equation}
M_{2}^{2}\in \lbrack 3,4]~\mathrm{GeV}^{2},\ s_{0}^{\prime }\in \lbrack
11,12]~\mathrm{GeV}^{2}.  \label{eq:Wind5}
\end{equation}%
The strong coupling $g_{4}$ is determined at the mass shell of the $\phi $
meson, i.e., at $q^{2}=m_{\phi }^{2}$.

Our computations yield%
\begin{equation}
g_{4}\equiv \mathcal{G}_{4}(-m_{\phi }^{2})=(6.7\pm 1.2)\times 10^{-1}\
\mathrm{GeV}^{-1}.  \label{eq:coup4}
\end{equation}%
The parameters of the fit function are $\mathcal{G}_{4}^{0}=0.66~\mathrm{GeV}%
^{-1}$, $c_{4}^{1}=-0.09$, and $c_{4}^{2}=-0.02$.

The width of the decay $\mathcal{M}\rightarrow J/\psi \phi $ is equal to%
\begin{equation}
\Gamma \left[ \mathcal{M}\rightarrow J/\psi \phi \right] =(2.6\pm 0.6)~%
\mathrm{MeV}.  \label{eq:DW4}
\end{equation}%
Then, it is not difficult to find the full width of $\mathcal{M}$
\begin{equation}
\Gamma _{\mathcal{M}}=(62\pm 12)~\mathrm{MeV}.  \label{eq:FullW}
\end{equation}%
The width of the hadronic molecule $\mathcal{M}$, within errors of
calculations and measurements, agrees with the LHCb datum from Eq.\ (\ref%
{eq:DataA}).


\section{Discussion and summing up}

\label{sec:Conclusion}

In this work, we have calculated the mass and width of the scalar molecule $%
\mathcal{M}=D_{s}^{+}D_{s}^{-}$ in the framework of the QCD sum rule
methods. The mass of $\mathcal{M}$ has been evaluated using the two-point
sum rule approach. The full width of $\mathcal{M}$ has been computed by
taking into account decay modes $\mathcal{M}\rightarrow D_{s}^{+}D_{s}^{-}$,
$\mathcal{M}\rightarrow \eta _{c}\eta ^{(\prime )}$, and $\mathcal{M}%
\rightarrow J/\psi \phi $. Strong couplings $g_{i}$ that determine the width
of these decays have been found in the framework of the three-point sum rule
method.

Our result for the mass $m=(4117\pm 85)~\mathrm{MeV}$ of the molecule $%
\mathcal{M}$ exceeds considerably the corresponding LHCb datum $m_{1\mathrm{%
exp}}$, but is consistent with $m_{2\mathrm{exp}}$. It is evident that $%
\mathcal{M}$ is significantly heavier than the resonance $X(3960)$, which
makes problematic its interpretation as $X(3960)$. The full width $\Gamma _{%
\mathcal{M}}=(62\pm 12)~\mathrm{MeV}$ of $\mathcal{M}$ is consistent with
the LHCb measurement $\Gamma _{2\mathrm{exp}}$ as well.

The resonance $X(3960)$ was examined in our article \cite{Agaev:2022pis} as
the tetraquark $X=[cs][\overline{c}\overline{s}]$ with quantum numbers $J^{%
\mathrm{PC}}=0^{++}$. We obtained the following predictions for the
parameters of $X$: the mass $m=(3976\pm 85)~\mathrm{MeV}$ and the width $%
\Gamma _{\mathrm{X}}=(42.2\pm 8.3)~\mathrm{MeV}$. The parameters of the
diquark-antidiquark state $X$ are in nice agreements with the LHCb data
given by Eq.\ (\ref{eq:Data}), therefore in Ref.\ \cite{Agaev:2022pis}, it
was identified with the resonance $X(3960)$.

In both the molecule and diquark-antidiquark pictures dominant decay modes
of the scalar four-quark meson $\overline{c}c\overline{s}s$ are channels $%
\mathcal{M}\rightarrow D_{s}^{+}D_{s}^{-}$ and $X\rightarrow
D_{s}^{+}D_{s}^{-}$, respectively. Decays of molecular-type resonances to
constituent mesons by falling apart are, naturally, preferable channels for
such states. The interpolating current for $\mathcal{M}$ given by Eq.\ (\ref%
{eq:C1}) has an explicitly $D_{s}^{+}D_{s}^{-}$ type structure. Therefore,
it couples mainly to the physical mesons $D_{s}^{+}$ and $D_{s}^{-}$. But
the current $J(x)$ couples also to other two-meson states. Formally, this
can be demonstrated using Fierz transformations of \ $J(x)$. To this end, it
is convenient to rewrite it in the following form
\begin{equation}
J=\delta _{am}\delta _{bn}\left[ \overline{s}_{a}i\gamma _{5}c_{m}\right] %
\left[ \overline{c}_{b}i\gamma _{5}s_{n}\right] .
\end{equation}%
After Fierz transformation it contains the following components%
\begin{eqnarray}
&&J(x)=\frac{\delta _{am}\delta _{bn}}{4}\left\{ \left[ \overline{s}%
_{a}\gamma _{\mu }s_{n}\right] \left[ \overline{c}_{b}\gamma ^{\mu }c_{m}%
\right] -\left[ \overline{s}_{a}s_{n}\right] \left[ \overline{c}_{b}c_{m}%
\right] \right.   \notag \\
&&\left. +\left[ \overline{s}_{a}i\gamma _{5}s_{n}\right] \left[ \overline{c}%
_{b}i\gamma _{5}c_{m}\right] +\cdots \right\} ,  \label{eq:C3}
\end{eqnarray}%
where the ellipses stand for axial-vector and tensor structures. Then,
rearranging the color indices by means of the equality
\begin{equation}
\delta _{am}\delta _{bn}=\delta _{an}\delta _{bm}+\epsilon _{abk}\epsilon
_{mnk},
\end{equation}%
with $\epsilon _{ijk}$ being the Levi-Civita epsilon, we find
\begin{eqnarray}
&&J(x)=\frac{1}{4}\left\{ \left[ \overline{s}_{a}\gamma _{\mu }s_{a}\right] %
\left[ \overline{c}_{b}\gamma ^{\mu }c_{b}\right] -\left[ \overline{s}%
_{a}s_{a}\right] \left[ \overline{c}_{b}c_{b}\right] \right.   \notag \\
&&\left. +\left[ \overline{s}_{a}i\gamma _{5}s_{a}\right] \left[ \overline{c}%
_{b}i\gamma _{5}c_{b}\right] +\cdots \right\} .
\end{eqnarray}%
The terms above are $\mathrm{S-S,V-V}$, and $\mathrm{PS-PS}$ type
interpolating currents that couple to relevant meson pairs. For example, $%
\mathrm{V-V}$ and $\mathrm{PS-PS}$ currents couple to $J/\psi \phi $, $%
J/\psi \omega $, $\psi ^{\prime }\phi $ and $\eta _{c}\eta ^{(\prime )}$
meson pairs (this list can be extended), respectively. Relative significance
of the $J(x)$ current's components can be seen by comparing strong couplings
at vertices $\mathcal{M}D_{s}^{+}D_{s}^{-}$, $\mathcal{M}\eta _{c}\eta
^{(\prime )}$, $\mathcal{M}J/\psi \phi $, etc., because only relevant
components of $J(x)$ contribute to three-point correlators. From a chain of
inequalities $g_{1}>g_{4}>g_{2}>g_{3}$, it is clear that $J(x)$ couples
dominantly to $D_{s}^{+}D_{s}^{-}$ and $J/\psi \phi $ mesons. Smallness of
the partial width $\Gamma \left[ \mathcal{M}\rightarrow J/\psi \phi \right] $
is connected with parameters (masses, decay constants) and quantum numbers $%
J^{\mathrm{PC}}=1^{--}$ of the final-state mesons.

In its turn, a diquark-antidiquark current can be expressed in terms of
molecule currents \cite{Chen:2022sbf,Xin:2021wcr}. Now, comparing strong
couplings of the diquark-antidiquark state $X=[cs][\overline{c}\overline{s}]$
with mesons $D_{s}^{+}D_{s}^{-}$, and $\eta _{c}\eta ^{(\prime )}$, we see
that $G>g_{1}>g_{2}$ \cite{Agaev:2022pis}. In general, one might explore the
vertex $XJ/\psi \omega $ and evaluate corresponding coupling, but the
contribution of the decay $X\rightarrow J/\psi \omega $ [$X\rightarrow
J/\psi \phi $ is forbidden kinematically] to $\Gamma _{\mathrm{X}}$ would be
negligible. Summing up, we can state that decays to $D_{s}^{+}D_{s}^{-}$ are
dominant channels for both the diquark-antidiquark structure $X$ and the
hadronic molecule $\mathcal{M}$: The resonances $X(3960)$ and $X_{0}(4140)$
were discovered in the $D_{s}^{+}D_{s}^{-}$\ mass distribution.

In the context of the sum rule approach the molecule $D_{s}^{+}D_{s}^{-}$
was also studied in Ref.\ \cite{Xin:2022bzt}. In accordance with this paper,
the mass of such hadronic molecule is equal to $(3980\pm 100)~\mathrm{MeV}$
and agrees with the LHCb data. It is worth noting that the authors did not
analyze quantitatively the width of this state. Our results for parameters
of $\mathcal{M}$, even within existing errors of calculations, does not
support molecule assignment for $X(3960)$.

By taking into account predictions for the mass $m$ and full width $\Gamma _{%
\mathcal{M}}$ of the molecule $\mathcal{M}=D_{s}^{+}D_{s}^{-}$ obtained in
the present work, we argue that the molecule $\mathcal{M}$ may be a
candidate to the structure $X_{0}(4140)$.

\end{document}